\documentclass[12pt]{article}
\usepackage{amsmath,amssymb,bm,graphicx}
\usepackage{color} 
\usepackage{hyperref}
\usepackage{slashed}
\usepackage{braket}
\usepackage{caption}
\usepackage{subcaption}
\usepackage{physics}

\usepackage{cite}
\numberwithin{equation}{section}

\begin{document}


\begin{center}
{\Large{\bf On Massive Neutrinos and Coherence \\
\vskip 0.3cm
in Neutrino Oscillations\footnote{Invited contribution to a special issue of Nuclear Physics B: ``Clarifying Common Misconceptions in High Energy Physics and Cosmology''.}}}
\end{center}
\vskip .3 truecm
\begin{center}
{\bf { Anca Tureanu}}
\end{center}

\begin{center}
\vspace*{0.4cm} 
{\it Department of Physics, University of Helsinki,\\ 
and\\
\vskip 0.1cm
Helsinki Institute of Physics,\\ P.O.Box 64, 
FI-00014 University of Helsinki,
Finland
}
\end{center}
\vspace*{0.2cm} 
\begin{abstract}
 We examine the central tenet of the current standard theory of neutrino oscillations, namely the assumption that neutrinos are emitted and detected as flavour neutrino states, which are coherent superpositions of massive neutrino states of different masses. We prove that all the quantum mechanical and quantum field theoretical arguments, including the invocation of the uncertainty principle and the wave packet description of massive neutrinos, entail the production of neutrinos as statistical ensembles of massive neutrino states. As the states in a statistical ensemble do not interfere, neutrino oscillations cannot be explained by the superposition of massive states. We point out that neutrino oscillations in vacuum can be consistently formulated in theories which include, among other assumptions, the premise that the asymptotic states are massless flavour neutrinos.

\end{abstract}

\section{Introduction}\label{sec:intro}

The current standard theory of neutrino oscillations \cite{Mohapatra, Fukugita, Bilenky, Giunti, Roulet, Valle, Xing} is predicated on the assumption that there exist flavour neutrino states $|\nu_\ell\rangle,\ \ell=e,\mu,\tau$, which are {\it coherent superpositions} of orthonormal massive neutrino states $|\nu_k\rangle,\ k=1,2,3$ with different kinematical masses $m_k,\ k=1,2,3$. We ignore here the hypothetical existence of extra species of sterile neutrinos. The formula that represents mathematically the above assumption is:
\begin{eqnarray}\label{states_mix}
| \nu_{\ell} \rangle
=
\sum_{k} U_{\ell k}^* \, | \nu_{k} \rangle
\,,
\end{eqnarray}
where $ U_{\ell k}$ are the elements of the Pontecorvo--Maki--Nakagawa--Sakata (PMNS) matrix, which is the mixing matrix for the left-handed flavour neutrino fields:
\begin{equation}
\nu_{\ell L}(x)
=
\sum_{k} U_{\ell k} \, \nu_{kL}(x),
\qquad
\ell=e,\mu,\tau
\,.
\label{fields_mix}
\end{equation}
The second assumption is that neutrinos are created and absorbed as flavour neutrinos and propagate in vacuum as massive neutrinos. The latter interfere and lead to an oscillatory flavour probability, which is known as neutrino oscillations \cite{Mohapatra, Fukugita, Bilenky, Giunti, Roulet, Valle, Xing}. The probability of transition from the flavour state $|\nu_\ell\rangle$ to the state $|\nu_{\ell'}\rangle$ for neutrinos of energy $E$, over a distance of propagation $L$, is given by the formula
\begin{equation}
P_{\nu_{\ell}\to\nu_{\ell'}}(L,E)
=
\sum_{k,j}
U_{{\ell}k}^{*}
\,
U_{{\ell'}k}
\,
U_{{\ell}j}
\,
U_{{\ell'}j}^{*}
\,
\exp\left( - i \, \frac{\Delta{m}^{2}_{kj} L}{2E} \right)
\,,
\label{P_osc}
\end{equation}
where $\Delta{m}^{2}_{kj}=m^2_k-m^2_j$.

Our purpose in this note is to examine the physical possibility of creating coherent superpositions of massive neutrino states by weak interaction processes. The definition of flavour neutrino states as in \eqref{states_mix} is an {\it ad hoc} expedient to formally achieve the interference of massive states and the flavour probability oscillation, in analogy with the oscillations of a two-level system in quantum mechanics. The actual preparation of coherent superpositions of states is, in general, not a trivial matter \cite{Nielsen, Rieffel, Wilde}. Can we establish that there exists a physical scheme by which superpositions of massive on-shell neutrinos can be created and interact? In the following sections, we shall demonstrate that the answer to this question is negative.

When discussing neutrino interactions, it is generally acknowledged that neutrinos are created and annihilated in weak interactions as probabilistic mixtures of massive neutrino states. All calculations of decay rates and cross sections in which the masses of neutrinos are kept explicitly are based on this concept \cite{Shrock}. On the other hand, when discussing neutrino oscillations, it is generally affirmed that neutrinos are created and annihilated as coherent superpositions of massive neutrino states \cite{Mohapatra, Fukugita, Giunti, Bilenky, Valle, Xing, Roulet}. These two statements are in stark conflict: a probabilistic mixture of some states is essentially different from a coherent superposition of the same states.

When we say that {\it we, the observers, prepare neutrinos in a certain initial state for an oscillation experiment}, that is not entirely true. However, we can prepare states of other particles that, by their decay, create neutrinos. In the most famous experiments that established neutrino oscillations and adiabatic matter conversion, the observer has not even this limited control over the source. So, it is Nature, not the observer, that "prepares" the neutrino states. If neutrinos have kinematic masses, by the rules of QFT, Nature delivers probabilistic mixtures of massive neutrinos. For the neutrinos to oscillate, after production, a statistical ensemble (the probabilistic mixture) must transform into a pure state (the coherent superposition), without the intervention of the observer. This is an implausible scenario and, in a nutshell, the reason why neutrino oscillations cannot be consistently explained in the currently standard theory.

The subject has a long history and there are already many dedicated monographs. Books distill and capture the essence of the field at the moment of writing. There is hardly any disagreement between the various authoritative monographs on the issue of the coherence of massive states, so we shall consider the selection of books \cite{Mohapatra, Fukugita, Giunti, Bilenky, Valle, Xing, Roulet} as general references and will avoid referring to research articles, which are too many to be properly and fairly cited. Nevertheless, it should be mentioned that there are quite a few alternative theoretical proposals for mechanisms of neutrino oscillations in vacuum; see, for example, the reviews \cite{Dolgov,Beuthe, Akhmedov_subtleties}. 

The paper is organized as follows: in Sec. \ref{sec:standard approach} we present the current standard approach to the coherent superposition of massive states, with the justifications that are generally agreed upon in books and articles. We selected three essential aspects of the problem, namely: i) the derivation of coherent superpositions of states from the $S$-matrix in Sec. \ref{sec:Smatrix}; 2) the compatibility between the coherent superposition of states and the incoherent summation of probabilities in the calculation of decay rates and cross sections for flavour neutrino processes in Sec. \ref{sec:interactions} and 3) the uncertainty principle argument and the wave-packet framework for the justification of coherence in Sec. \ref{sec:uncertainty_wp}. In Sec. \ref{sec:failure of coherence} we analyze the previously presented arguments and show that they are in contradiction with the principles of quantum field theory and quantum mechanics. The analysis invariably leads to the conclusion that massive neutrinos are produced in weak interactions as statistical ensembles and cannot form coherent superpositions. In Sec. \ref{sec:outlook} we consider other perspectives, reflecting on the theory of neutrino oscillations in a wider context.

\section{Standard approach to flavour neutrino states, coherence and oscillations}\label{sec:standard approach}

\subsection{$S$-matrix and the flavour neutrino states }\label{sec:Smatrix}

The justification for the assumption of coherence of massive states as in \eqref{states_mix} is rarely found in the literature. To our knowledge, there is only one line of argument that is presented in detail in Ref. \cite{Giunti} and partly in Ref. \cite{Bilenky}. The argument goes as follows.

Neutrinos are usually produced in decay processes, and their flavour is identified by the accompanying charged lepton in the charged-current interaction. Generically, a decay can be written as
\begin{equation}
\text{P}_{\text{i}} \to \text{P}_{\text{f}} + \ell^{+} + \nu_{\ell},
\label{decay}
\end{equation}
where $\text{P}_{\text{i}}$ is the initial decaying particle
and $\text{P}_{\text{f}}$ represents all other final particles except the ones specified. The particle $\text{P}_{\text{i}}$ may have other decay channels in addition to \eqref{decay}. 

The notation $\nu_\ell$ in \eqref{decay} encodes any neutrino that accompanies the charged antilepton $\ell^+$, according to the charged-current interaction in the leptonic sector:
\begin{eqnarray}\label{LCC}
{\cal L}_{CC}&=&-\frac{g}{\sqrt2}\bar\ell_L(x)\gamma_{\mu} {\nu}_{\ell L}(x)W^\mu(x)+h.c.\cr
&=&-\frac{g}{\sqrt2}\sum_{\ell,k}\bar\ell_L(x)U_{\ell k}\gamma_{\mu} {\nu}_{k L}(x)W^\mu(x)+h.c.,
\end{eqnarray}
where ${\nu}_{\ell L}(x)$ are the flavour neutrino fields and ${\nu}_{k L}(x)$ are the massive neutrino fields with mass $m_k$, respectively. In effect, the expression \eqref{LCC} in terms of massive neutrino fields implies that the reaction \eqref{decay} is taking place through three different channels (or more, if sterile neutrinos are included):
\begin{equation}
\text{P}_{\text{i}} \to \text{P}_{\text{f}} + \ell^{+} + \nu_{k},\ \ \ k=1,2,3.
\label{decay'}
\end{equation}

The decay, as well as any particle scattering, is described by a scattering matrix $\mathsf{S}$. The initial particle, upon decay, results in one configuration of an infinite set of possible final configurations. By configuration we mean a final state $|f\rangle$, specified by asymptotic particle states with their well-defined energies and momenta. Sometimes, the evolution of the initial state $|i\rangle=|\Phi(-\infty)\rangle$  in which the system was prepared long before the interaction took place, at $T=-\infty$, is written as 
\begin{equation}
|\Phi(\infty)\rangle
=
\mathsf{S} \, | i \rangle
,
\label{final state}
\end{equation}
where $|\Phi(\infty)\rangle$ encodes all possible final configurations at $t=\infty$. The transition amplitude for a specific final state $|f\rangle$ is given by the corresponding element of the $S$--matrix, namely
\begin{equation}
\langle f|\Phi(\infty)\rangle
=
\langle f| \, \mathsf{S} \, |i \rangle =S_{fi}.
\label{matrix element}
\end{equation}
The state $|\Phi(\infty)\rangle$ can then be expanded in a complete set of states in the asymptotic Hilbert space, which is a direct sum of all the Fock spaces of the individual fields that appear in the Lagrangian. The expansion reads:
\begin{equation}
|\Phi(\infty)\rangle
=
\sum_f|f\rangle\langle f| \, \mathsf{S} \, |i \rangle =\sum_f S_{fi}|f\rangle.
\label{final expansion}
\end{equation}
The matrix elements $S_{fi}$ between well defined initial and final states are complex functions of the masses and momenta of those states. Consequently, the expression \eqref{final expansion} appears formally as a coherent superposition, with prescribed coefficients and fixed relative phases, of all the possible final states into which the state $|i\rangle$ can evolve. (We will see later, in Sec. \ref{sec:S_matrix_fallacy} that this interpretation is actually wrong.)

Applying this line of argument to the neutrino production by the reactions \eqref{decay'}, one obtains:
\begin{equation}
| \Phi(\infty) \rangle
=
\sum_{k} \mathcal{A}^{\text{P}}_{\ell k} \, | \nu_{k} , \ell^{+} , \text{P}_{\text{f}} \rangle
+
\ldots
\,,
\label{nu final state}
\end{equation}
where the dots represent other decay channels, if they exist, and
\begin{equation}
\mathcal{A}^{\text{P}}_{\ell k}
=
\langle \nu_{k} , \ell^{+} , \text{P}_{\text{f}} | \Phi(\infty) \rangle
=
\langle \nu_{k} , \ell^{+} , \text{P}_{\text{f}} |
\,
\mathsf{S}
\,
| \text{P}_{\text{i}} \rangle
\,.
\label{nu elements}
\end{equation}
Naturally, the expansion coefficients $\mathcal{A}^{\text{P}}_{\ell k}$ depend on the masses and momenta of all initial and final particle states, which is omitted in this way of writing. They also depend linearly on the elements of the PMNS matrix, i.e.
\begin{equation}\label{coeff}
\mathcal{A}^{\text{P}}_{\ell k}=U_{\ell k}^{*}
\,
\mathcal{M}^{\text{P}}_{\ell k},
\end{equation}
where $\mathcal{M}^{\text{P}}_{\ell k}$ is a kinematical factor. This factor, in the limit of ultrarelativistic massive neutrinos of energy $E\gg m_k$, is well approximated by a common value, which is actually the Standard Model expression for massless neutrinos of the same energy \cite{Bilenky}: 
\begin{equation}\label{approx}\mathcal{M}^{\text{P}}_{\ell k}\approx \mathcal{M}^{\text{P}(SM)}_{\ell}.\end{equation}

One may go one step further and project $|\ell^{+} , \text{P}_{\text{f}} \rangle$ out of the final state and normalize the result, thus defining a {\it production flavour neutrino state} \cite{Giunti, Bilenky} with prescribed coefficients, inherited from \eqref{nu final state}:
\begin{equation}
| \nu_{\ell}^{\text{P}} \rangle
=
\sum_{k}
\frac{\mathcal{M}^{\text{P}}_{\ell k}}{ \sqrt{ \sum_{j} |U_{\ell j}|^{2} \, |\mathcal{M}^{\text{P}}_{\ell j}|^{2} } }
\,
U_{\ell k}^{*}
\,
| \nu_{k} \rangle.
\label{flavour production}
\end{equation}
The superscript ${\text{P}}$ indicates that this flavour state is defined for the specific decay \eqref{decay}, which fixes the coefficients $\mathcal{A}^{\text{P}}_{\ell k}$. When the condition \eqref{approx} is fulfilled, one recovers the standard expression for the flavour states \eqref{states_mix}.

For neutrino detection processes, one defines by a similar procedure a phenomenological {\it detection flavour neutrino state} \cite{Giunti}. The detection usually takes place via a scattering of the type
\begin{equation}\label{detection}
\nu_{\ell} + \text{D}_{\text{i}} \to \text{D}_{\text{f}} + \ell^{-},
\end{equation}
where $\text{D}_{\text{i}}$ is the target and $\text{D}_{\text{f}}$ represents other final particles besides the charged lepton. The detection flavour neutrino state would read
\begin{equation}
    | \nu_{\ell}^{\text{D}} \rangle
=
\sum_{k}
\frac{\mathcal{M}^{\text{D}}_{\ell k}}{ \sqrt{ \sum_{j} |U_{\ell j}|^{2} \, |\mathcal{M}^{\text{D}}_{\ell j}|^{2} } }
\,
U_{\ell k}^{*}
\,
| \nu_{k} \rangle,\label{flavour detection}
\end{equation}
where
\begin{equation}\label{detection coeff}
\mathcal{A}^{\text{D}}_{\ell k}=\mathcal{M}^{\text{D}}_{\ell k}U_{\ell k}^{*}=\langle \nu_{k} , \text{D}_{\text{i}} | \mathsf{S}^{\dagger} \, | \text{D}_{\text{f}} , \ell^{-} \rangle
\,.
\end{equation}
In general, the kinematical factors for production and detection are naturally different, therefore there is a slight mismatch between the coefficients in \eqref{flavour production} and \eqref{flavour detection}.

For ultrarelativistic neutrinos, using \eqref{approx}, a similar formula for the detection states, and the unitarity of the PMNS matrix, $\displaystyle \sum_{k} |U_{\alpha k}|^{2} = 1$, both \eqref{flavour production} and \eqref{flavour detection} lead to the standard flavour neutrino states \eqref{states_mix}. This is true in all the present oscillation experiments, which are not sensitive to the differences in the kinematical factors for the various massive neutrinos.

The formulas \eqref{flavour production} and \eqref{flavour detection} display mathematically the coherence of massive neutrino states, a feature that can be traced down to the general formula \eqref{final expansion}. This is the only mathematical derivation that we know about for the superposition with prescribed coefficients arising in the production or detection of neutrinos. 
Using these formulas one can recover the neutrino oscillation probability formulas, see, e.g., \cite{Giunti}. The amplitude of the oscillations will be slightly modified due to the kinematical factors, but the patterns of oscillations given by the phase differences will remain unchanged.

In Sec. \ref{sec:S_matrix_fallacy} we shall break down this argument and show that scattering theory does not support the idea that a coherent superposition of all possible final states may physically occur. In effect, we shall prove that massive neutrinos with different masses can be produced only as statistical mixtures.

\subsection{Coherence vs incoherence in neutrino interactions}\label{sec:interactions}

Let us return to the main features of the current standard approach to neutrino oscillations. The formula \eqref{nu final state}, as it stands, literally implies coherence between all possible final states of the scattering. However, the canon of quantum field theory holds that each decaying particle results in {\it only one of the possible final states}, and not in a coherent superposition of all of them. This is generally acknowledged also for the production of neutrinos: the decay rates and scattering cross sections are given by incoherent sums over the different final state configuration, meaning also the different channels corresponding to the different massive neutrinos \cite{Shrock}. Massive neutrinos are believed to be the physical particles that propagate in space and time with definite kinematical properties (mass, momentum, energy) and, by this token, they are the asymptotic neutrino states.

We are thus in front of a controversy: the decays lead to incoherent massive neutrino production; nevertheless, the emitted neutrinos propagate as coherent superpositions of massive neutrinos. We postpone the examination of this controversy to Sec. \ref{sec:incompatibility}. Here we shall present an argument \cite{Giunti} that is meant to show that this controversy is immaterial. It goes as follows: in view of the definition of the flavor state \eqref{flavour production}, and taking into account the amplitudes \eqref{nu elements},
the amplitude of the decay process \eqref{decay}
with flavour neutrino final state is formally written as:
\begin{equation}
\mathcal{A}^{\text{P}}_{\ell}
=
\langle \nu_{\ell}^{\text{P}} , \ell^{+} , \text{P}_{\text{f}} |
\,
\mathsf{S}
\,
| \text{P}_{\text{i}} \rangle
=
\left( \sum_{i} |\mathcal{A}^{\text{P}}_{\ell i}|^{2} \right)^{-1/2}
\sum_{k} \mathcal{A}^{\text{P}*}_{\ell k}
\,
\langle \nu_{k} , \ell^{+} , \text{P}_{\text{f}} |
\mathsf{S}
| \text{P}_{\text{i}} \rangle
=
\sqrt{ \sum_{i} |\mathcal{A}^{\text{P}}_{\ell i}|^{2} }
\,.
\label{flavour_amplitude}
\end{equation}
Consequently,
the decay probability is given by the incoherent
sum of the probabilities of production of the different massive neutrinos,
\begin{equation}
|\mathcal{A}^{\text{P}}_{\ell}|^{2}
=
\sum_{i} |\mathcal{A}^{\text{P}}_{\ell i}|^{2}=\sum_{i} 
|\langle \nu_{i} , \ell^{+} , \text{P}_{\text{f}} |
\mathsf{S}
| \text{P}_{\text{i}} \rangle|^2
\,.
\label{flavour_probability}
\end{equation}
It is deduced from here \cite{Giunti} that the coherence of the flavour neutrino states has no bearing on the derivation of the decay probability, which can be calculated by using either flavour states or an incoherent mixture of massive states.  The conclusion is that, although neutrinos propagate as coherent superpositions of massive neutrinos, they are however emitted as statistical (incoherent) mixtures of the massive neutrinos, as prescribed by quantum field theory.

The same result is predicted for the detection of incoming flavour states. We will show in Sec. \ref{sec:incompatibility} that this is not the case and, actually, the use of coherent flavour states is incompatible with the quantum field theory calculations for detection processes.

\subsection{Uncertainty principle and wave packets}\label{sec:uncertainty_wp}

The argument above may be considered sufficient to solve the controversy, but it leaves us still with an ambiguity: if the analysis of decay cannot distinguish coherent from incoherent, then what is the rationale for considering that neutrinos are created, after all, as coherent superpositions of massive states?

The answer is given by invoking the uncertainty principle \cite{Fukugita, Giunti, Bilenky, Xing}. Since the analysis of oscillations involves localized sources and detectors and the distance between them, this means that the neutrinos have to be localized, therefore they have to be described by wave packets and not plane waves. The localization of the produced neutrinos implies that the uncertainty in the position is finite and fixed by the characteristic dimensions of the source. By the uncertainty principle, this in turn means that the spread in momentum of the neutrino wave packet cannot be less than a certain value. If the spread in momentum is larger than the mass differences between any of the massive states, it can be argued that one cannot possibly know which of the massive states was produced. It is then commonly concluded that, since one ignores which massive neutrino was produced, one has to include all the possibilities in a {\it coherent superposition} of massive neutrino wave packets \cite{Fukugita, Giunti, Bilenky, Xing}. Why coherent superposition and not probabilistic mixture? The literature does not answer this question, but we shall analyze it in Sec. \ref{sec:wp}.

What are then the coefficients of the superposition of wave trains? The uncertainty principle argument cannot provide a quantitative solution to this problem. It is again commonly agreed upon that the coefficients are exactly the same as for the coherent superposition of plane waves, cf. eq. \eqref{flavour production}
and \eqref{flavour detection}. Ignoring the kinematical factors, which are indeed negligible for ultrarelativistic neutrinos, the flavour neutrino state evolution in the wave packet approach is written as \cite{Fukugita,Giunti, Xing}:
\begin{equation}\label{wavepacket}
|\nu_\ell(x,t)\rangle=\left[\frac{1}{2\pi\sigma_x^2}\right]^{1/4}\sum_k U_{\ell k}^*\int d^3 p\,e^{\left[i(p_k x-E_k t)-\frac{(x-v_kt)^2}{4\sigma_s^2}\right]}|\nu_k(p)\rangle,
\end{equation}
where $\sigma_x$ is the configuration space width of the wave packet, $p_k$ and $E_k$ are the average momentum and energy of the massive state $k$, $v_k=\partial E_k/\partial p_k=p_k/E_k$ is the group velocity of the wave train. Different wave packet descriptions have to be used for the source and the detector. Then, after a number of plausible approximations, the oscillation probability in vacuum can be recovered in the simple standard form \cite{Fukugita,Giunti, Xing}.

In this framework, the current standard description of neutrino oscillations goes as follows: the flavour neutrino is produced as the coherent superposition of massive neutrino wave trains. The wave trains propagate with slightly different group velocities because of their different masses. While the wave packets overlap during propagation, the massive states interfere and produce oscillations according to the formula \eqref{P_osc} in the ultrarelativistic regime. If allowed to propagate over very long distances, the wave trains separate so much that their interaction with the detector particles does not produce interference any longer \cite{Valle}. When the wave packets are separated, no oscillation is possible. There is no interference if
two wave packets are separated by more than the width of each packet \cite{Fukugita}. However, the interaction of neutrinos takes place whether they are in a coherent superposition or not. As a result, neutrinos can be detected as flavour neutrinos or as massive neutrinos (see, for example, the solar neutrinos).

\begin{figure}[h]
    \centering
    \includegraphics[width=16cm]{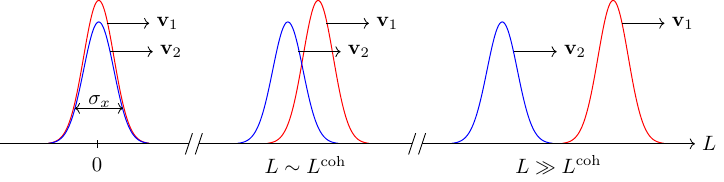}
        \caption{\small Standard representation of wave packets separation for two-neutrino mixing (reproduced from Ref. \cite{Giunti}).}
    \label{fig:separation}
\end{figure}

The traveled distance at which the separation of wave packets becomes significant is called coherence length and in this formalism is given by the expression:
\begin{equation}
L_{coh}=\frac{4\sqrt 2 E^2}{|\Delta m^2_{kj}|}\sigma_x.
\end{equation}
For sensible choices of $\sigma_x$ and assuming the energies of neutrinos in the oscillation experiments, the coherence length is very large and the experiments are not affected by decoherence due to wave packet separation. The separation of the wave packets during propagation is often depicted as in Fig. \ref{fig:separation}.

In Sec. \ref{sec:wp} we will show that the argument of the uncertainty principle and the use of wave packets do not support the interpretation that flavour neutrinos are created as coherent superpositions massive neutrinos. The invocation of wave packets provides, on the contrary, one of the most powerful arguments in favour of mixed state production of neutrinos. 

\section{The failure of coherence}\label{sec:failure of coherence}

\subsection{The fallacy of the $S$-matrix definition of flavour states}\label{sec:S_matrix_fallacy}

As we emphasized earlier, the only mathematical derivation of flavour states created or destroyed in weak interactions, with prescribed coefficients for the massive states, is the one presented in Sec. \ref{sec:Smatrix}. The argument is premised on the existence of an all-encompassing final state $|\Phi(\infty)\rangle$ given by \eqref{final state}, which is a coherent superposition of all possible final-state configurations that can arise in the scattering of initial-state particles, according to \eqref{final expansion}:
\begin{equation*}
|\Phi(\infty)\rangle
=
\sum_f|f\rangle\langle f| \, \mathsf{S} \, |i \rangle =\sum_f S_{fi}|f\rangle.
\end{equation*}
This description is nevertheless misleading. The problem is that it assumes a physical significance of {\it all possible final-state configurations that occur together}. In a coherent superposition like \eqref{final expansion}, the symbol $|\Phi(\infty)\rangle$ stands for a "state" that is different from all states $|f\rangle$, and is supposed to represent the actual configuration after scattering. But that is not the case.

The correct, experiment-based way of thinking about the "all-encompassing final state" of a scattering is as a {\it mixed state}, described by a density matrix, in which each of the possible final states appears with its own probability: 
\begin{equation}
\rho(\infty)
=
\sum_f|S_{fi}|^2|f\rangle\langle f| .
\label{final density matrix}
\end{equation}
Scattering in quantum field theory means the evolution of an initial state $|i\rangle$, which is an ensemble of free particles, into a given final state $|f\rangle$, with a probability given by the square of the corresponding $\mathsf{S}$-matrix element, $|S_{fi}^2|$. The $\mathsf{S}$-matrix is defined by its matrix elements, $\langle f|\mathsf{S}|i\rangle$, which enable us to find the probabilities of individual transitions. The density matrix $\rho({\infty})$ is diagonal and its rank is infinite\footnote{In contrast, the rank of a density matrix describing a pure state is always equal to 1.}.

In experiments, the initial state is prepared as an ensemble of free particles with definite quantum numbers and kinematic properties, at mascroscopic distances from one another. They approach each other and interact in a microscopical region. The products of interaction are an ensemble of free particles, also with well-defined characteristics, which fly apart at macroscopic distances. The final state is an eigenstate of the free Hamiltonian. The number of possible final states is continuous infinity. In a given scattering, only one of the final states is produced and no other. We do not know {\it a priori} which final state will be realized, and the process is governed by probabilities. Our ignorance of the final state realized is encoded in the statistical mixture \eqref{final density matrix}.

The fact that all possible final states form a statistical mixture and not a coherent superposition is encoded in the well-known fact that the cross sections and decay widths, differential or total, are calculated as a sum over the probabilities of all final states. This is sometimes named an incoherent sum. But an incoherent sum in quantum physics is a sum over {\it possible} outcomes, not over actually realized outcomes: in one single scattering, only one of the possible outcomes is realized; in a large number of identically prepared samples (initial states), different outcomes will occur, with a specific probability, for each decay or scattering instance. All this is well known and under no debate in particle physics; therefore, it has to be applied equally to processes involving neutrinos.

Returning to the production of neutrinos in the light of the above discussion, we can re-examine the decay \eqref{decay} comparatively with \eqref{decay'}, i.e.
\begin{equation}
\text{P}_{\text{i}} \to \text{P}_{\text{f}} + \ell^{+} + \nu_{\ell}\ \ \ \ \ \ \ \mbox{vs}\ \ \ \ \ \ \ \text{P}_{\text{i}} \to \text{P}_{\text{f}} + \ell^{+} + \nu_{k},\ k=1,2,3.
\label{decay''}
\end{equation}
The Hamiltonian of interaction 
\begin{equation}\label{H_CC}
{\cal H}^{CC}(x)
=
\frac{g}{\sqrt2}\sum_{k}
U_{\ell k}^{*}
\,
\overline{\nu_{k}}(x)
\,
\gamma^{\mu}
\left( 1 - \gamma^{5} \right)
\ell(x)W_\mu(x)
\end{equation}
describes only the latter version of the decay, in which each individual final state contains one of the massive neutrinos $\nu_k$. 

Suppose that we consider the decay
\begin{equation}
\pi^+\to \mu^++\nu_\mu,
\end{equation}
which is in effect realized through three different channels:
\begin{equation}
\pi^+\to \mu^++\nu_k,\ \ k=1,2,3.
\end{equation}
When one given pion decays, the final state with muons can be $|\mu^+,\nu_1\rangle$, or $|\mu^+,\nu_2\rangle$, or $|\mu^+,\nu_3\rangle$. We do not know which one is realized, but we know that the ratio of probabilities is $|U_{\mu 1}|^2:|U_{\mu 2}|^2:|U_{\mu 3}|^2$, assuming that the differences between the kinematical factors are negligible for ultrarelativistic neutrinos. As far as neutrinos are concerned, the final state is a statistical mixture described by the ensemble of pure states 
\begin{equation}\left\{{\cal P}_k=|U_{\mu k}|^2,|\nu_k\rangle\right\}_{k=1,2,3},\label{ensemble}\end{equation} 
where ${\cal P}_k$ denotes the probability of the state $|\nu_k\rangle$ in the mixture.
In other words,
\begin{equation}
|\nu_\mu\rangle =|\nu_k\rangle, \ \ \ \mbox{with probability}\ |U_{\mu k}|^2,\ k=1,2,3.
\end{equation}
The same ensemble is expressed as a diagonal density matrix of rank bigger than 1:
\begin{equation}
\rho(\nu_\mu)
=
\sum_k|U_{\mu k}|^2|\nu_k\rangle\langle\nu_k | .
\label{neutrino density matrix}
\end{equation}

The language of pure vs. mixed states offers yet another perspective on the "purity" of flavour states. We ignore for the moment the previous arguments about the impossibility of producing coherently massive neutrinos. If the muon neutrino were indeed a pure state itself, i.e. the coherent superposition of pure mass states:
\begin{equation}
| \nu_{\mu} \rangle
=
\sum_{k} U_{\mu k}^* \, | \nu_{k} \rangle
\,,
\end{equation}
the density matrix would be rank 1 and nondiagonal:
\begin{equation}
\rho(\nu_\mu)_{pure}
=
\sum_{k,j}U^*_{\mu k}U_{\mu j}|\nu_k\rangle\langle\nu_j | .
\label{pure neutrino density matrix}
\end{equation}
The nondiagonal terms, called coherences, quantify the observability of the relative phase between the pure massive states. It is well known in quantum mechanics that not all quantum-mechanical states can be superposed coherently, due to selection or superselection rules. Coherent superposition of the states $|\nu_k\rangle$ and $|\nu_j\rangle$ in vacuum is possible only if there exists a self-adjoint operator $\hat{\cal O}$ associated with a physical observable, such that
\begin{equation}
\langle \nu_k|\hat{\cal O}|\nu_j\rangle \neq 0.
\end{equation}
The transition from one on-shell massive neutrino to another can happen through weak interactions, but with involvement of other on-shell particles. All these cannot occur when the neutrinos propagate in vacuum; therefore, no operator $\hat{\cal O}$ with the above property exists. In other words, neutrino states of different masses belong to different superselection sectors. As a result, in vacuum, the superposition principle for massive neutrinos is inhibited \cite{Strocchi}. This is another way of understanding on quantum mechanical grounds why flavour states cannot exist as pure states in vacuum, even if we pay no attention to the mechanism of production of neutrinos. 

 A final contradiction with quantum mechanics is the following: In vacuum, massive neutrinos can be considered {\it closed quantum systems}, because they propagate without interaction. Their coherent superpositions as flavour states are also regarded as non-interacting. In a closed quantum system, a pure state, under unitary evolution, remains pure. Decoherence does not occur. Consequently, a pure flavour state should not decohere. This is in contradiction with the decoherence of massive states by propagation over long distances in vacuum, represented in Fig. \ref{fig:separation}.

To conclude, the $\mathsf{S}$-matrix rationale does not support the creation of flavour neutrino states as coherent superpositions of massive states. If the asymptotic states are massive neutrinos, they can be produced and detected only as statistical mixtures.

\subsection{The fallacy of coherent states compatible with incoherent production and detection}\label{sec:incompatibility}

In Sec. \ref{sec:interactions} was presented an argument that, if the flavour neutrinos are described as coherent superpositions of massive states \eqref{flavour production}, the probability of the neutrino production is the same as what is expected when the massive states are produced incoherently \eqref{flavour_probability}. This claim is also extended \cite{Giunti} to detection processes. Here, we will show that the formal use of coherent detection flavour states \eqref{flavour detection} actually does not lead to the same scattering probability as the incoherent summation over different channels corresponding to different massive neutrinos.

Why is this aspect important? Let us assume, as in Sec. \ref{sec:interactions}, that neutrinos are produced and detected as coherent flavour states and they oscillate between the source and the detector. In this scenario, to extract experimentally the parameters of neutrino oscillations, one relies on the fact that the number of transitions as a function of the distance $L$ between the source and the detector and the energy $E$ of the neutrinos is given by
\begin{equation}
N_{\ell\ell'} (L)\propto \Gamma_\ell P_{\nu_\ell\to\nu_{\ell'}} (L)\sigma_{\ell'},
\end{equation}
where $\Gamma_\ell$ is the decay rate of the source particles with emission of $\nu_\ell$ and $\sigma_{\ell'}$ is the cross section of the scattering by which the neutrino $\nu_{\ell'}$ is detected. The number of transitions depends also on the energy of the neutrino, but we can omit this in our discussion. The production rate $\Gamma_\ell$ and the detection cross section $\sigma_{\ell'}$ are measured in independent experiments. For ultrarelativistic neutrinos, the measured values are compatible with the Standard Model calculations using massless neutrinos. Nevertheless, for the consistency of the procedure, {\it it is important that the calculations using coherent superpositions of massive neutrinos in either the final or initial states match the Standard Model calculations when the ultrarelativistic limit is taken.}

Now, let us return to the main issue of this section. The derivation of the detection flavour neutrino states \cite{Giunti} is even less convincing than the derivation of the production states, but we will not stop to discuss those intricacies. We take at face value the formula \eqref{flavour detection} with \eqref{detection coeff}. Then, the transition amplitude for a given coherent flavour neutrino in the initial state is found to be:
\begin{equation}\label{detection TA}
\mathcal{A}^{\text{D}}_{\ell}=\langle\text{D}_{\text{f}} , \ell^{-}   | \mathsf{S} \, | \nu_{\ell}^{\text{D}} , \text{D}_{\text{i}} \rangle=\left( \sum_{i} |\mathcal{A}^{\text{D}}_{\ell i}|^{2} \right)^{-1/2}
\sum_{k} \mathcal{A}^{\text{D}}_{\ell k}
\,\langle\text{D}_{\text{f}} , \ell^{-}   | \mathsf{S} \, | \nu_{k} , \text{D}_{\text{i}} \rangle=\sqrt{ \sum_{i} |\mathcal{A}^{\text{D}}_{\ell i}|^{2} }
\,,
\end{equation}
therefore,
\begin{equation}
|\mathcal{A}^{\text{D}}_{\ell}|^2= \sum_{i} |\mathcal{A}^{\text{D}}_{\ell i}|^{2} 
\,
\end{equation}
For ultrarelativistic neutrinos, this formula is well approximated by the SM model result:
\begin{equation}\label{scattering coherent}
|\mathcal{A}^{\text{D}}_{\ell}|^2= \sum_{i}|U_{li}|^2 |\mathcal{A}^{\text{D}(SM)}_{\ell}|^{2} =|\mathcal{A}^{\text{D}(SM)}_{\ell}|^{2}
\,,
\end{equation}
where $\mathcal{A}^{\text{D}(SM)}_{\ell}$ is the transition amplitude of the process \eqref{detection}, calculated in the Standard Model with massless flavour neutrinos of the same energy.

This result appears to say that the scattering probability for a {\it flavour neutrino in an initial state} is equal to the {\it sum of the probabilities} for each massive neutrino in the coherent superposition.

However, this is {\it not} what we expect to obtain if the initial state is a mixture of particles with different masses. Recall that in such a case, one has to {\it average over the possible initial states}, that is, to multiply the square of the transition amplitude for a given initial state by the probability of finding that state in the mixture.

Let us see what happens when we calculate the scattering probability according to the prescriptions of quantum field theory. The source emits $|\nu_{\ell'}\rangle$ and in the detector is observed the charged lepton $\ell$. The charged current interaction that occurs in the detector is described by the hermitian conjugates of the Hamiltonian terms \eqref{H_CC}. We assume that $|\nu_{\ell'}\rangle$ emitted at the source propagates as the coherent superposition of mass eigenstates given by \eqref{states_mix}. For definiteness, we assume also equal-momenta ultrarelativistic massive neutrinos. At a distance $L$ from the source, where the detector is placed, the state will be:
\begin{eqnarray}\label{state at  L}
|\nu(L, E)\rangle=\sum_{k} U_{\ell' k}^* \, e^{-i(m_k)^2\frac{L}{E}}\, | \nu_{k} \rangle
\,. 
\end{eqnarray}
The detector "sees" the incoming neutrinos as a statistical mixture of $|\nu_k\rangle,\ k=1,2,3$, with the probabilities for each given by the squares of the coefficients in \eqref{state at  L}:
\begin{equation}\label{detection mixture}
\{{\cal P}_k=|U_{{\ell'} k}|^2,|\nu_k\rangle\}_{k=1,2,3}.
\end{equation}
The scattering probability of each individual massive neutrino in the detector is proportional to
\begin{equation}
|\langle\text{D}_{\text{f}} , \ell^{-}   | \mathsf{S} \, | \nu_{k} , \text{D}_{\text{i}} \rangle|^2=|U_{lk}|^2|\mathcal{A}^{\text{D}(SM)}_{\ell}|^2\,.\end{equation}
To compose the total detection cross section, in view of the fact that the initial neutrino is one of the possible $|\nu_k\rangle$, with the probability given by \eqref{detection mixture}, we have to {\it average over the initial neutrino states}, which in this case means
\begin{equation}\label{scattering incoherent}
|\mathcal{A}^{\text{D}}_{\ell}|^2_{\text{QFT}}=\sum_k |U_{{\ell'} k}|^2 |U_{lk}|^2|\mathcal{A}^{\text{D}(SM)}_{\ell}|^2<|\mathcal{A}^{\text{D}(SM)}_{\ell}|^2.
\end{equation}
Thus, the correct calculation of the incoherent probability of absorption of the massive states \eqref{scattering incoherent} differs by the factor $\sum_k |U_{{\ell'} k}|^2 |U_{lk}|^2$ from the calculation of the absorption probability of the formal coherent superpositions of massive states
\eqref{scattering coherent}. Note that in calculating \eqref{scattering incoherent}, we even assumed that the neutrino had oscillated according to \eqref{state at  L} before being detected. However, the probability that the neutrino emitted as $\nu_{\ell'}$ is detected as $\nu_{\ell}$ is the same at any distance between the source and the detector, because the probability of finding $\nu_k$ in the mixture remains constant during propagation and equal to $|U_{{\ell'} k}|^2$. The detector is oblivious to the presumed coherence of the incoming state and oscillation is not detected.

We have proven that the mechanism of interaction of neutrinos as statistical mixtures, which is agreed upon in the literature, is in conflict with the formal definition of the interaction of coherent superpositions of states. This confirms once more the inconsistency in the definition of coherent flavour states.

\subsection{The fallacy of the wave packets superposition solution}\label{sec:wp}

Most books and research articles dealing with neutrino oscillations do not delve into the actual scheme by which flavour neutrinos arise as coherent superpositions of massive neutrinos.
The usual rationale for coherence is based on the uncertainty principle, which we have already reviewed in the beginning of Sec. \ref{sec:uncertainty_wp}; in short, runs the argument, because of localization of emission and uncertainty principle, we cannot know which massive state was produced, therefore it {\it must} be a {\it coherent superposition} of all the massive states \cite{Giunti, Fukugita, Xing}. 

This line of argument implicitly acknowledges that {\it one} of the massive states is objectively produced in the decay of a certain particle, but we simply ignore which one. {\it However, the lack of knowledge about the state of a system means, by definition, that the "state" is a statistical mixture.} The coherent superposition, on the other hand, would imply that we have perfect knowledge about the state \cite{Nielsen, Rieffel, Wilde} (though, of course, this does not mean perfect knowledge about the outcome of any experiment done with that state, but this is just how quantum mechanics works). Thus, the correct premise that the uncertainty principle does not allow us to know which massive state was emitted leads logically to the conclusion that the emitted "entity" was not a coherent superposition, but a statistical mixture of massive states. The uncertainty principle does not breed coherence.

Neutrinos are particles; therefore, we cannot detect or even represent "half-a-neutrino" or a "part-of-a-neutrino". Elementary particles that are absolutely stable are described, according to Wigner's classification, in terms of their mass and spin \cite{BoLoTo}. Real particles are represented as Fock space states. This is compatible with the corpuscular interpretation of quantum field theory and the widely used notion of particle in the theory of scattering. {\it A quantum mechanical particle is a packet of energy and momentum that is not composed of smaller packets} \cite{Zwiebach}. When we describe particles by wave packets, one wave packet represents one particle, with definite mass and spin.

Wave packets of massive neutrinos are brought into play because the energy and momentum of neutrinos cannot be known with sufficient precision to discriminate between the various massive states, and the source and detector are localized. This is in fact valid for any particle observed in an experiment. The whole theory of scattering, in its most rigorous form, must be developed in terms of wave packets of the in- and out-states \cite{Coleman, Peskin, Weinberg I}. The scattering theory is predicated on the fact that the initial/final particles are far apart from each other in space in the far past/future, therefore noninteracting. This implies a necessary notion of localization, and therefore wave packets, for the in- and out-states. A derivation of the scattering cross section using wave packets to represent particles can be found in Ref. \cite{Peskin}. Nevertheless, it turns out that the wave packets' shapes and spreads drop out completely from the calculation of the cross sections, mainly because the detectors cannot resolve positions at the level of de Broglie wavelength, nor the small variations of momenta around the central value of the wave packet. The same is true in the derivation of the LSZ reduction formula. Whether the incoming and outgoing particles are represented by wave packets or by plane waves, the corpuscular interpretation of the theory of scattering is the same; consequently, only one of the possible final configurations is objectively achieved in any individual scattering or decay. In a process like \eqref{decay}, {\it one single neutrino is produced for each decaying particle, which means one single neutrino wave packet in the final state}.

Let us return to the mental image represented in Fig. \ref{fig:separation}. In the wave-packet framework, the corpuscular interpretation is even more expressive: the wave trains describing two particles with different masses and different velocities overlap significantly during the propagation over long distances. This means that, necessarily, {\it the two wave packets exist simultaneously at the same space points}. Moreover, for interference to occur, the packets have to be coherent with each other, in accord with formula \eqref{wavepacket}. 

We shall examine the plausibility of this description by running the following {\it Gedankenexperiment}: Assume that a muon neutrino in a two-neutrino mixing scheme is produced by pion decay at $t=0, x=0$ and it propagates undisturbed over a distance $L$ much longer than the coherence length, such that the two wave packets corresponding to the two massive neutrinos $\nu_1$ and $\nu_2$ are well separated\footnote{This is not an entirely academic exercise: the C$\nu$B neutrinos, if ever observed, would fulfill this condition.}. Recall that the {\it measurement} of a particle in QFT annihilates the particle\footnote{In contrast, in quantum mechanics, we can measure a certain property of a particle, for example its spin polarization in a certain direction, without destroying the particle itself.}. At the distance $L$ from the source, we place a detector. The fastest of the two massive neutrinos, say $\nu_1$, is detected/annihilated. We want to ask ourselves what is, in theory, the probability to detect the particle $\nu_2$ that is trailing behind.

\begin{figure}[h]
    \centering
    \includegraphics[width=10cm]{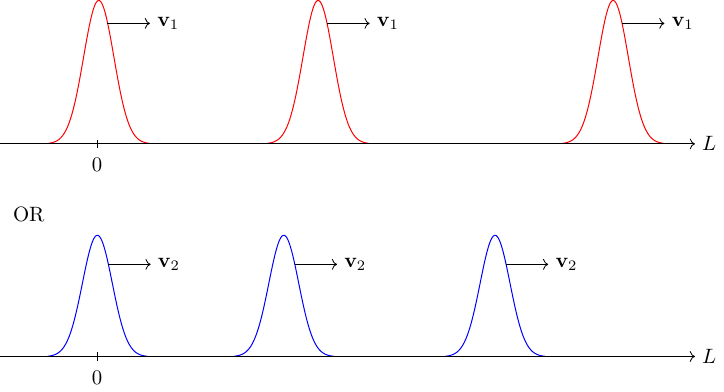}
        \caption{\small Massive neutrinos are produced as mixed states in weak interactions. The wave packets of two massive neutrinos composing one flavour state cannot overlap, coherently or incoherently, because they do not exist simultaneously.}
    \label{fig:separate wp}
\end{figure}

There are two possibilities:
\begin{enumerate}

\item The probability of detecting $\nu_2$ is not equal to zero, i.e. $\nu_2$ can interact after $\nu_1$ had been detected. This means that one single pion decayed into one muon and {\it two} neutrinos; in other words, the flux of neutrinos would double (or triple, in the case of three-neutrino mixing). This scenario is excluded.

\item The probability of detecting the particle $\nu_2$, after the particle $\nu_1$ had been detected, is theoretically zero. This can be interpreted in two ways:

\begin{enumerate}
\item the particle $\nu_2$ disappeared at the same time when the particle $\nu_1$ was annihilated in the detection process. Yet particle $\nu_2$ was far behind particle $\nu_1$, at least "far" compared to the range of weak interactions. Real particles propagating in vacuum do not disappear, unless they interact. If particle $\nu_2$ disappeared, it means that it had interacted; therefore its probability of being detected is not zero. Thus, we come to a contradiction with the original hypothesis that the probability of detecting particle $\nu_2$ is zero. This scenario is also excluded.

\item the particle $\nu_2$ did not exist/was not produced in the first place. Only particle $\nu_1$ had been produced, but we did not know which of the two was emitted. This means that only one wave packet (the red one, with group velocity $\bf{v}_1$ in Fig. \ref{fig:separation}) propagated throughout the distance $L$. Consequently, there was {\it never} overlap and the detection experiment would have given the same result no matter how close the detector was placed to the source. This scenario is the only possible one according to the rules of quantum field theory and again tells us that the emitted neutrinos are mixed states of massive particles, as represented in Fig. \ref{fig:separate wp}. 
\end{enumerate}
\end{enumerate}

Thus, the uncertainty principle argument is consistent with the quantum field theoretical paradigm, entailing that the neutrino states created and destroyed in weak interactions are probabilistic mixtures of massive neutrino states. Accordingly, the graphic description of massive neutrino wave packets propagating from source to detector is rather the one in Fig. \ref{fig:separate wp} and not the one in Fig. \ref{fig:separation}.

\section{Outlook}\label{sec:outlook}

We have analyzed, in the light of quantum field theory and quantum mechanics principles, widely used and frequently repeated arguments for the production and detection of flavour neutrino states as coherent superpositions of massive on-shell states. We came to the conclusion that the premises of those arguments lead instead to the conclusion that neutrinos are produced and detected as statistical mixtures of massive states. The formally defined flavour neutrino states \eqref{states_mix} cannot be physically realized. The experimentally observed neutrino oscillations in vacuum necessitate a different theoretical framework. Although we have not touched upon the neutrinos in matter, the conclusion is not so drastic in that case: the adiabatic conversion of solar neutrinos, for example, can be derived, with similar results, using statistical mixtures of massive neutrinos.

Various drawbacks of the current standard neutrino oscillation theory have been known for a long time and alternative theoretical frameworks have been proposed. It is not our purpose here to review the alternatives, many of which can be found in the references \cite{Dolgov,Beuthe, Akhmedov_subtleties} and the curated trove of neutrino literature on the site \href{https://www.nu.to.infn.it/}{Neutrino Unbound}. There is hardly any consensus at the moment on a consistent theory to replace the standard one, and the alternative approaches do not get sufficiently well studied. Most alternative frameworks maintain the existence of neutrinos with different kinematic masses; recently, the idea that neutrinos may have only refractive masses has also been explored. Wolfenstein suggested that neutrino oscillations and conversion in matter can be formulated in analogy with birefringence \cite{Wolfenstein1, Wolfenstein2}, realizing that one can detect the coherent interference between the forward scattered neutrino and the incident one. Recently, we proposed a new theory of vacuum oscillations, extending the birefringence analogy to include the neutrino Yukawa interaction with the Brout--Englert--Higgs (BEH) vacuum \cite{ONT}. In this scenario, neutrinos are created and annihilated in weak interactions as massless flavour particles, but the vacuum acts similarly to an optically active medium, providing refractive mass and flavour conversion. Although the neutrinos propagate in vacuum with a velocity less than the speed of light, nonetheless the group velocity is universal for a given energy, irrespective of the flavour composition of the wave. Coherence is maintained indefinitely in vacuum and contradictions with quantum field theoretical and quantum mechanical principles do not arise.

Neutrino oscillations in vacuum can be formulated even in the absence of interactions with the BEH field, assuming that the massless Standard Model neutrinos have Yukawa-type interactions with dark matter \cite{Choi, AYS_refractive, Ge}. 

Neutrinos have captivated the interest of particle physicists mainly because they are supposed to have very tiny masses, out of the range of common expectation. Their oscillations in vacuum, on the other hand, are considered by many as a simple quantum mechanical process analogous to a basic two-level system. Nevertheless, it may well be that neutrinos do not have kinematic masses at all, while the oscillations are an exotic phenomenon, worthy of more reflection. It remains for the experiments to discern what type of masses, kinematical or refractive, the neutrinos have. Hopefully, this article brings clarity about the limitations and deficiencies of the current standard theory of neutrino oscillations and will stimulate the debate and creative solutions for the particle oscillation theory. 

\section*{Acknowledgments}

I am grateful to Masud Chaichian, Esko Keski-Vakkuri, Markku Oksanen and, in particular, to Dieter Haidt, for illuminating discussions and comments on the manuscript.

\end{document}